\pdfoutput=1
\documentclass[fullpage]{article}
\usepackage{graphicx}
\usepackage{amsmath}
\usepackage{afterpage}

\usepackage{tikz} 
\usepackage{caption}
\usepackage{subfigure}
\usepackage{multirow}
\usepackage{xcolor}

\usepackage{listings}

\usepackage{placeins}

\usepackage{enumerate}

\usepackage[title]{appendix}

\usepackage{tablefootnote}

\usepackage{refcount}

\usetikzlibrary{arrows}
\usetikzlibrary{shapes}
\usetikzlibrary{shapes.multipart}

\tikzset{
    *|/.style={
        to path={
            (perpendicular cs: horizontal line through={(\tikztostart)},
                                 vertical line through={(\tikztotarget)})
            -- (\tikztotarget) \tikztonodes
        }
    }
}

\tikzset{every picture/.style={line width=1pt}}

\definecolor{columbiablue}{rgb}{0.61, 0.87, 1.0}
\definecolor{classicrose}{rgb}{0.98, 0.8, 0.91}
\definecolor{lightgray}{rgb}{0.83, 0.83, 0.83} 
\definecolor{inchworm}{rgb}{0.7, 0.93, 0.36}
\definecolor{backcolor}{rgb}{0.95,0.95,0.92}

\lstdefinestyle{mystyle}{
    backgroundcolor=\color{backcolor},
    basicstyle=\ttfamily\footnotesize,
    breakatwhitespace=false,
    breaklines=true,
    captionpos=b,
    keepspaces=true,
    showspaces=false,
    showstringspaces=false,
    showtabs=false,
    tabsize=4
}

\newcommand{\tikzxmark}{%
\tikz[scale=0.23] {
    \draw[line width=0.7,line cap=round] (0,0) to [bend left=6] (1,1);
    \draw[line width=0.7,line cap=round] (0.2,0.95) to [bend right=3] (0.8,0.05);
}}
\newcommand{\tikzcmark}{%
\tikz[scale=0.23] {
    \draw[line width=0.7,line cap=round] (0.25,0) to [bend left=10] (1,1);
    \draw[line width=0.8,line cap=round] (0,0.35) to [bend right=1] (0.23,0);
}}

\begin{document}

\title{A case study on different one-factor Cheyette models for short maturity caplet calibration}
\author{Arun Kumar Polala, Bernhard Hientzsch}
\maketitle

\begin{abstract}
In \cite{PDML}, we calibrated a one-factor Cheyette SLV model with a local volatility that is linear in the benchmark forward
rate and an uncorrelated CIR stochastic variance to 3M caplets of various maturities. While caplet smiles for many maturities
could be reasonably well calibrated across the range of strikes, for instance the 1Y maturity could not be calibrated 
well across that entire range of strikes. Here, we study whether models with alternative local volatility terms and/or
alternative stochastic volatility or variance models can calibrate the 1Y caplet smile better across the strike range
better than the model studied in \cite{PDML}. This is made possible and feasible by the generic simulation, pricing, and
calibration frameworks introduced in \cite{PDML} and some new frameworks presented in this paper. We find that some 
model settings calibrate well to the 1Y smile across the strike range under study. In particular, a model setting with a local volatility 
that is piece-wise linear in the benchmark forward rate together with an uncorrelated CIR stochastic variance and one with 
a local volatility that is linear in the benchmark rate together with a correlated lognormal stochastic volatility 
with quadratic drift (QDLNSV) as in \cite{LNSVSeppetal_2023} calibrate well. We discuss why the later might be a preferable model.      
\end{abstract}

\section{Introduction}

\begin{figure}[h]
    \centering
    \includegraphics[width=0.60 \linewidth]{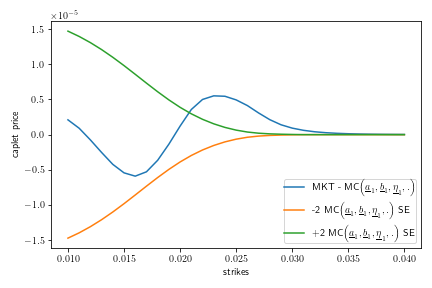}
    \caption{Calibrating the 1yr Maturity Caplet smile, Cheyette Model with Linear Benchmark 
    Forward Rate Local Volatility and CIR SV. Figure and data from \cite{PDML}. 
    Shown is difference between model price (MC) and market price (MKT), MC error bars are shown for comparison. }
    \label{pdmlpaper1yrmaturityexampleI}
\end{figure}

In \cite{PDML}, we proposed and tested a new calibration methodology on the example of 3M caplets of various maturities,
calibrating one-factor Cheyette models with a local volatility that is linear in the 3M forward rate and 
an uncorrelated CIR stochastic variance. While most maturities could be calibrated reasonably well across a range
of strikes, the 1Y maturity could not be calibrated well across the entire range of those strikes. 
When calibrating the wings reasonably well, strikes in the range of about 200bps to 300bps -- right of ATM -- 
could not be calibrated well, see Figure \ref{pdmlpaper1yrmaturityexampleI}. 

Here, we study whether one-factor Cheyette models with different forms of local volatilities and/or 
stochastic volatilities or variances will calibrate better. We find that some model settings calibrate well, 
in particular, a model setting with a local volatility 
that is piece-wise linear in the benchmark forward rate together with an uncorrelated CIR stochastic variance
and one with a local volatility that is linear in the benchmark rate together with a correlated 
lognormal stochastic volatility with quadratic drift (QDLNSV) as in \cite{LNSVSeppetal_2023} lead to good calibrations.  
We see that the presented generic simulation, pricing, and calibration frameworks make such calibration and 
modeling studies, feasible, efficient, and even easy and can serve as a good foundation for modeling. 

The paper is organized as follows: In section \ref{OFCM}, we introduce the Cheyette models under consideration, including 
motivation for and references for the various forms.   
In section \ref{CapletsFloorlets}, we introduce caplets and floorlets and how they can be priced in the Cheyette model
by simulation. We then introduce the Generic Simulation Framework in section \ref{GenSimFW} which allows us to simulate 
various payoffs under various models, including under varying parameters, in a convenient and efficient fashion. 
In section \ref{MCPCG}, we present how the Generic Simulation Framework can be used to code generate 
MC pricers for various backends (TensorFlow, NumPy, JAX) and how calibration can be implemented by optimization around such pricers. 
In section \ref{PDMLIntro}, we discuss how the recently proposed parametric differential machine learning (PDML) 
methodology can be used for such parametric pricing and calibration.
In section \ref{whichway}, we quickly discuss when to apply code generation versus PDML approaches for the parametric pricers.
Finally, in section \ref{Results}, we apply 
these pricing and calibration methods to calibrate Cheyette models in various model settings, inspect the results,
and identify the best models for our purposes. In section \ref{Conclusion}, we conclude. 

\section{One factor Cheyette Models}\label{OFCM}
The dynamics of a single factor Cheyette model with states $\left\{x_t, y_t \right\}$, volatility process $\sigma_{t}$, 
and constant mean reversion is given by
\begin{eqnarray}
dx_t &=& \mu_{t} dt + \sigma_{t} dW_t, \\
dy_t &=& \left(\sigma_{t}^{2} - 2\lambda y_t \right)dt, 
\end{eqnarray}
where $\lambda$ is mean reversion speed and $W_t$ is a standard Brownian motion. $x_t$ is the diffusive Cheyette factor
while $y_t$ is a locally deterministic auxiliary factor ensuring that the curve stays arbitrage free. 
The drift term $\mu_{t}$ is given by
\begin{equation}
\begin{split}
\mu_{t}& =  \left(y_t - \lambda x_t \right) \quad  \text{(risk neutral measure)}, \\
            & = \left(y_t - \lambda x_t  - G(T-t) \sigma_{t}^{2} \right) \quad \text{(T-Fwd measure)},
\end{split}
\end{equation}
where $G(x) = \frac{1 - h(x)}{\lambda}$ and $h(x) = e^{-\lambda x}$. In this paper,
we use $\lambda=0.03$, a common value from the literature. The choice of $\lambda$ does not seem 
to impact the results too much beyond numerical stability. Thus, we are using a standard, stable value.
The volatility process $\sigma_{t}$ is specified as follows:
\begin{equation}
\begin{split}
\sigma_{t}& = \sigma(t, x_t, y_t) \quad \text{(no SV)}, \\
                & = \sigma(t, x_t, y_t)  \vartheta_{t} \quad \text{(SV)},   
\end{split}
\end{equation}
where $\vartheta_{t}$ is some stochastic process. 
In this paper, we consider the following functional forms for the local volatility function $\sigma(t,x_t,y_t)$:
\begin{enumerate}[(i)]

\item Linear short rate volatility (`LinSRLV'): 
\begin{equation}\label{shortratevol}
\sigma\left(t,x_t,y_t\right) = a + b \times \left( f(0,t) + x_t \right),
\end{equation}
where $f(0,t)$ is the initial forward rate. This form has been used, for instance, by \cite{hoorens2011cheyette}, both without stochastic volatility 
or variance and with uncorrelated CIR stochastic variance. 
Short(est) maturity rates can be observed in the market or extracted from discount curves which are
constructed and saved at least every business day in financial institutions and can be analyzed over time. 
It is typically assumed that the historical evolution of risk factors and the risk neutral evolution as 
implied from traded options are equivalent with a measure change involving various 
market prices of risk which change the drift but not the volatility during the measure change. 
Hence, observed behaviors of the volatility of the short rate or of another rate can inform the 
structure of the volatility term. The volatility of the observed short rate seems to be linear in 
the short rate within certain regimes.  

\item Linear benchmark forward rate volatility (`LinBRLV'): 
\begin{equation}\label{benchmarkratevol}
\sigma\left(t,x_t,y_t\right) = a + b \times f\left(t, t+ \delta \right),
\end{equation}
where $f\left(t, t + \delta \right)$ is the forward rate for tenor $\delta$ and can be computed as follows: 
\begin{equation}
f\left(t , t + \delta \right)  = f(0,t + \delta) + h(\delta) \left(x_t + y_t G(\delta) \right).
\end{equation}
This form is the ``standard form" advocated for and used by \cite{andersenpiterbarginterest} and many in industry and academia, 
both without stochastic volatility or variance and with uncorrelated CIR stochastic variance.

\item Volatility linear in Cheyette factor (`LinXLV'):
\begin{equation}\label{localvol}
\sigma\left(t,x_t,y_t\right) = a + b \times x_t.
\end{equation}
This form has been used and advocated for in \cite{chibane2012explicit}.

\item Piece-wise linear benchmark forward rate volatility (`PwLinBRLV'): 
\begin{equation}\label{benchmarkPL}
\begin{split}
\sigma\left(t,x_t,y_t\right)& =  a_1 1_{f(t,t + \delta) < K_1} + \sum_{i=1}^{n-1} \left( a_i + b_i \left(f(t,t + \delta) - K_i \right) \right) 1_{K_i \leq f(t,t + \delta) < K_{i+1}} \\
                                     & \quad + a_n 1_{K_n \leq f(t,t+\delta)},
\end{split}
\end{equation}
where $\{K_i \}_{i = 1}^{i = n}$ denotes the set of strikes in increasing order, 
$b_i = \frac{a_{i+1} - a_i}{K_{i+1} - K_i},$ and $\{ a_i \}_{i = 1}^{i = n}$ denotes the constant terms. 

As discussed above, studying how the volatility of the rates behaves under the historical measure can inform the 
volatility models under pricing measure since the processes under two equivalent measures would share the same 
volatility term (but not drift). \cite{deguillaume2013nature} studied the volatility of interest rates in particular in dependence 
of the rate level and found a piece-wise-linear form to be a good approximation over a very long observation period 
in various rate regimes. Thus, local volatilities for benchmark instantaneous forward rates, piece-wise linear in said benchmark forward rates;
or local volatilities for short rates, piece-wise linear in short rates, would be supported by the studies in \cite{deguillaume2013nature}. 
\end{enumerate} 

For the stochastic volatility or variance (SV) dynamics, we consider two types of stochastic models: 
CIR stochastic variance and lognormal stochastic volatility with quadratic drift (`QDLNSV'). 
These two models and other SV models were compared in \cite{robustSVSeppetal_2023}. 
In this paper, we investigate CIR SV and lognormal SV with quadratic drift (QDLNSV) 
in the context of the Cheyette model. The stochastic processes $\vartheta_{t}$ for CIR SVs and 
lognormal SV with quadratic drift is given below:

\begin{itemize}
\item CIR stochastic variance (`CIRSV'): Here, $ \vartheta_{t} = \sqrt{z_t}$ with an SDE for $z_t$:
\begin{equation}
dz_t = \theta \left( z_0 - z_t \right)dt + \eta(t) \sqrt{z_t} dZ_t,
\end{equation}
where $\theta$ is the mean reversion speed of variance, $\eta(t)$ is volatility of variance, 
$z_0 = z(0) = 1$, and we assume $dZ_t dW_t \equiv 0$. 
We are using $\theta=0.2$, but we confirmed that the calibration results are not sensitive to the 
exact value of the mean reversion speed of the variance.  
For CIR SV models, we have the Feller condition which says that only if twice the product
of mean reversion speed ($\theta$) and long-term mean (here, $z_0=1$) is larger than the 
square of the volatility coefficient $\eta(t)$ is the process $z_t$ positive 
($2 \theta z_{\infty} \geq \eta^2$ with $z_\infty=z_0$ here). 
If the Feller condition is not satisfied, $z_t$ can be zero, can spend substantial time absorbed in or close to zero, 
and will behave deterministically in the first instant after hitting zero, contradicting for many what seems to be observed in the market.  
If either the local or the stochastic volatility term is zero, the volatility term will be zero,
and thus Girsanov transformations cannot change the drift, 
leading to additional modeling restrictions when modeling behavior both in the observational 
measure and pricing measure with a standard risk neutral pricing approach.  
In this paper, we enforce the Feller condition in the calibrations.

\item Correlated CIR SV (`CorCIRSV'): Here, still  $ \vartheta_{t} = \sqrt{z_t}$ with an SDE for $z_t$, but nonzero correlation is allowed.
Under risk-neutral measure, the SDE is given as:
\begin{equation}
    dz_t =  \left[ \theta \left( z_0 - z_t  \right)  \right] dt + \beta(t) \sqrt{z_t} dW_t + \epsilon(t) \sqrt{z_t} dW_t, \\
\end{equation}
while under $T$-forward measure it is given as:
\begin{equation}
dz_t =  \left[ \theta \left( z_0 - z_t  \right) - G(T-t) \sigma(t,x_t,y_t) \beta(t) z_t  \right] dt + \beta(t) \sqrt{z_t} dW_t + \epsilon(t) \sqrt{z_t} dW_t, \\
\end{equation}
where $\theta$ is the mean reversion speed of variance, $\eta(t)$ is volatility of variance, $\beta(t)=\rho \eta(t)$, 
$\epsilon(t)=\sqrt{1-\rho^2} \eta(t)$, and $z_0 = z(0) = 1$, and we assume $dZ_t dW_t \equiv \rho$. 
We are using $\theta=0.2$ as in the 
uncorrelated case above and the same Feller condition applies (and is enforced in the calibrations). 
As can be seen, non-zero correlation leads to additional drift terms under $T$-forward and annuity measures 
which means that the standard approximations leading to displaced Heston models do not apply, and the
approximate fast solvers proposed for the zero correlation case cannot be applied.

\item Lognormal stochastic volatility with quadratic drift (`QDLNSV'): 
Here, an SDE for $\vartheta_t$ is directly specified. This SDE under the two measures
is given as:
\begin{eqnarray}
\begin{split}
d\vartheta_t &= \left( \kappa_1 + \kappa_2 \vartheta_t \right) \left( \theta - \vartheta_t \right)dt + \beta(t) \vartheta_t dW_t + \epsilon(t) \vartheta_t dZ_t \quad \text{(under risk neutral)} \nonumber \\
d\vartheta_t &= \left( \left( \kappa_1 + \kappa_2 \vartheta_t \right) \times \left( \theta - \vartheta_t \right)  - G(T-t) \sigma(t,x_t,y_t) \beta(t) \vartheta_{t}^{2} \right)dt \\
                  & + \beta(t)\vartheta_t dW_t +  \epsilon(t) \vartheta_t dZ_t \quad \text{(under T-Fwd measure)},
\end{split}
\end{eqnarray}
where $\kappa_1$ and $\kappa_2$ are linear and quadratic mean reversion speeds, respectively, 
$\theta$ is the mean level of volatility, 
$\beta(t)$ and $\epsilon(t)$ are deterministic terms, and we assume $dZ_t dW_t \equiv 0$. 
We follow \cite{LNSVSeppetal_2023} in that we are choosing to assume $\kappa_1=\kappa_2=\kappa$ to avoid over-parameterization 
and pick a $\kappa$ that seems to lead to a calibrated volatility of volatility that is relatively stable across 
expiries, which seems to be around $\kappa=0.25$ for the USD market. We also follow \cite{LNSVSeppetal_2023} by picking $\lambda=0.025$.
As is common in rates and some applications in equity, we use the stochastic volatility only to shape the distribution but 
not to impact the expected volatility by choosing $\theta=1$ and $\vartheta_0=1$. 
Note that, in this case we have different dynamics for the stochastic volatility based on the choice of 
measure just as for correlated CIR SV, whereas in the case of 
uncorrelated CIR SV the dynamics remains the same irrespective of the measure. 
\end{itemize}

Besides previous work and studies on the volatility of rates in the observational measure, 
there is another important consideration in the literature 
when choosing a particular form of local and stochastic volatility or variance. 
Most papers argue that closed-form or almost closed-form pricers for calibration 
instruments are required to obtain a calibration that will be fast enough and thus forms are 
preferred that lead to settings in which such pricers
can be found or that can be approximated by settings in which such pricers can be or have been found. 
(Forward term rate) Caplets and floorlets can often be priced as 
options on zero-coupon bonds assuming that the underlying model has closed-form or 
quasi-closed-form solutions for such options. However, 
caplets and floorlets can also be seen as options on forward rates. 
The forward term rate under the forward measure corresponding to the payment time 
(which is equals to the end time of the forward rate) of the caplet is a martingale 
and the SDE for it can be derived and turns out to be structurally similar to 
or the same as the SDE for the Cheyette factors. The swap rate under the annuity 
measure corresponding to the fixed leg of the swap is a martingale
and the  SDE for it can be derived and turns out to be structurally similar to the 
SDE for the Cheyette factor. In the swaption case, a number of 
approximations and assumptions are necessary to approximate the swap rate SDE 
with an SDE of the same form with constant coefficients. Settings with 
constant, linear, and quadratic local volatility without stochastic volatility 
can be solved in (quasi-)closed-form; thus leading to people 
choosing such local volatility forms and using those solvers. 
The coefficients of the local volatility form for the swap rate might be time 
dependent and  mildly stochastic but can be approximately reduced by 
time-averaging to settings with constant coefficients.
Settings with constant or linear local volatility with uncorrelated CIR 
stochastic variance can be mapped to (displaced) Heston and 
Heston solvers (with direct integration or by FFT or other transform methods) 
can be used for pricing. While these (displaced) Heston 
solvers can treat correlation between the underlying and the stochastic variance, 
assuming such nonzero correlation between the stochastic variance
and the Cheyette factor does not allow the approximations and transformations 
that map the forward rate or swap rate SDE to a displaced Heston model. 
Theoretically, other uncorrelated stochastic volatility or variance models 
could be used in a similar fashion 
assuming that solvers are available and the appropriate forms of time-averaging 
have been derived, but it seems that 
no such time-averaging has been published, and such solvers have not been used so far. 

Alternatively, one can try to derive approximate solvers by various expansions, 
for instance as in \cite{LNSVSeppetal_2023}. This leads to similar
needs to derive approximations and expansions and to implement them; 
while possibly allowing settings in which the stochastic volatility 
can be correlated to the underlying factor and/or rate. 

We here price and calibrate based on MC simulation pricing and are thus not 
constrained by such considerations as long as the proposed pricing 
and calibration approaches are fast enough to be used. Since we are pricing 
under the original, not approximated, problem, we can price
arbitrarily well by using enough MC samples assuming the discretization 
is convergent. Thus, MC simulation is typically used to benchmark approximation 
approaches as in \cite{LNSVSeppetal_2023}. 
We can control accuracy for the calibration pricers by choosing appropriate
numbers of samples and other discretization parameters.

\section{Calibration Instruments: Caplets and Floorlets}\label{CapletsFloorlets}

(Forward term rate) Caplets and floorlets are European style interest rate derivatives. 
Let $T_{1} < T_{2}$, then the payoff of the caplet  and floorlet at $T_{2}$ for notional amount $N$ is
\begin{equation}\label{caplet/floorletpayoff}
N \delta^\omega(T_1,T_2) \left( \omega \left( F_{T_{1}}^{F}\left(T_{1},T_{2}\right) - K  \right)  \right)^+,
\end{equation}
where
\begin{equation}
\begin{split}
\omega& = 1 \quad \text{(caplet)} \\
           & = -1 \quad \text{(floorlet)},
\end{split}
\end{equation}
$K$ is the strike price, $F_{.}^{F}\left(T_{1},T_{2}\right)$ is the forward rate 
for the time period $\left[T_{1},T_{2}\right]$, $T_{1}$ is the reset date, 
$T_{2}$ is the payment date and $\delta^{\omega}$ is an appropriate day count 
fraction between $T_{1}$ and $T_{2}$ for the caplet and floorlet.

The forward rate can be computed from the discount factor for the forward/forecasting curve $P^F(t,T)$ 
- which represents the discount factor for that curve for $T$ as seen from $t$.
\begin{equation}
F_{t}^{F}\left(t^{S},t^{E}\right) = \frac{1}{\delta^F\left(t^{S},t^{E}\right)}\left(\frac{P^{F}\left(t, t^{S}\right)}{P^{F}\left(t, t^{E}\right)} - 1 \right),
\end{equation}
with $\delta^F$ being the appropriate day count fraction for the forecasting forward rate.

We consider the two-curve setting, in which the forecasting curve $P^F$ and the 
discounting curve $P^D$ can be different. Based on hypothesis (\textbf{S0}) 
as formalized and introduced in \cite{MH_multicurve}, we can write the forward rate based on the forecasting curve 
as an affine function of the forward rate based on the discounting curve. 
\begin{equation}
F_{t}^{F}\left(t^{S},t^{E}\right) = m F_{t}^{D}\left(t^{S}, t^{E} \right) + s,
\end{equation}
with 
\begin{eqnarray}
 m & = &  \beta_{0}^{F}\left(t^{S},t^{E} \right), \\
 s & = & \frac{\beta_{0}^{F}\left(t^{S},t^{E} \right) - 1}{\delta(t^S,t^E)}, \\
\beta_{0}^{F}(t^S,t^E)  &=& \frac{P^{F}(0,t^S)}{P^{F}(0,t^E)} \frac{P^{D}(0,t^E)}{P^{D}(0,t^S)}.
\end{eqnarray}
Here, we assume that the time conventions and day counts for the different curves are the same, 
i.e., $\delta^D$ = $\delta^F$ = $\delta$. 
Otherwise, the structure of the formulas stays the same 
while the exact form of the coefficients becomes somewhat more complicated. 
For further discussion of this setting and the above expressions, see \cite{PDML}.

For the Cheyette models under consideration, the discount factor functions $P^{.}(t,T)$ for the 
two curves (forecasting and discounting) are given as a function of state variables $(x_t, y_t)$. 
\begin{equation}\label{discountfactorfunc}
P^{.}(t,T;x_t,y_t) = \frac{P^{.}(0,T)}{P^{.}(0,t)} \exp \left( -G(T-t) x_t -\frac{1}{2} G^2(T-t) y_t \right). 
\end{equation}  
Note that the discount factor function $P^{.}$ depends only on $(x_t, y_t)$ even in the case of SV. 
Also, note that $P^{.}(0,T;x_0,y_0)$ is equal to the initial curve $P^{.}(0,T)$. 
Further, the discount factor function remains the same for both measures (T-Fwd and risk neutral), 
just that $x_t$ and $y_t$ are given by SDE systems with different drifts for the different measures.

The payoff of a caplet ($\omega=1$) and floorlet ($\omega=-1$) given in 
Equation \ref{caplet/floorletpayoff}  can be simplified to:  
\begin{equation}\label{twoasonepayoff}
N \delta \left( \omega \left(F_{T_{1}}^{F}\left(T_{1},T_{2}\right) - K \right)  \right)^+  = 
N \left( \omega \left( \frac{m}{P^{D}(T_{1},T_{2})} - \hat{K} \right) \right)^+ ,
\end{equation}
with
\begin{equation}
\hat{K} = 1 + K \delta . 
\end{equation}
Here, we assume that the day count for the caplet or floorlet $\left( \delta^{\omega} \right)$ 
is the same as the forecasting and discounting forward rate day count ($\delta$). 
Using the discount factor function given in Equation \ref{discountfactorfunc}, we can simplify 
\begin{equation}
\left( \frac{m}{P^{D}(T_{1},T_{2})} - \hat{K} \right) = \left( \frac{P^{F}(0,T_1)}{P^{F}(0,T_2)}  \exp \left( G(T_2-T_1) x_{T_1} + \frac{1}{2} G^2(T_2-T_1) y_{T_1} \right)  - \hat{K} \right)
\end{equation}
and write Equation \ref{twoasonepayoff} as:
\begin{equation}
N \left( \omega \left( \frac{m}{P^{D}(T_{1},T_{2})} - \hat{K} \right)  \right)^+  = 
N \left( \omega \left( p_F  \exp \left( c_x x_{T_1} + c_y y_{T_1} \right)  - \hat{K} \right)  \right)^+ , \label{twoasonepayoff1} 
\end{equation}
for appropriate coefficients $p_F$, $c_x$, and $c_y$. For further details, we refer to \cite{PDML}.

Finally, the price of a caplet and floorlet as of time $0$ is given by 
\begin{equation}\label{capletpriceexpectation}
\mathsf{Num}_{0} \times E \left[  \frac{N \left( \omega \left( p_F  \exp \left( c_x x_{T_1} + c_y y_{T_1} \right)  - \hat{K} \right) \right)^+}{\mathsf{Num}_{T_2}} \right], 
\end{equation}
with $\mathsf{Num}_t$ given by 
\begin{equation}
\begin{split}
\mathsf{Num}_t& = P^D(t,T) \quad \text{under  T-Fwd measure},\\
                       & = e^{\int_{0}^{t} r(u)du} \quad \text{under risk neutral measure}. 
\end{split}
\end{equation}
In case of forward measure, we chose $T_2$-Fwd measure in which case  $\mathsf{Num}_t$ = $P^{D}(t,T_2)$. 
As, $P^D(T_2,T_2) \equiv 1$ we have $\mathsf{Num}_{T_2} \equiv 1$ and $\mathsf{Num}_0 = P^D(0,T_2)$.

In One-Factor Cheyette models, we can write the short rate $r$ as follows:
\begin{equation}
r(t) = f(0,t) + x_t.
\end{equation}
So, under the risk neutral measure the num\'{e}raire $\mathsf{Num}_t$ can be expressed as follows:
\begin{eqnarray}
\mathsf{Num}_t &=& e^{\int_{0}^{t} \left( f(0,u) + x_u \right) du},\\
d\mathsf{Num}_t &=& \left(f(0,t) + x_t \right) \mathsf{Num}_{t} dt. \label{mmanumerairedyn}
\end{eqnarray}
Note that in this case, $\mathsf{Num}_0$ = 1. 
Based on Equation \ref{mmanumerairedyn}, we can jointly simulate $\mathsf{Num}_{t}$ along with the state variables $(x_t, y_t)$ 
(and volatility variable $\vartheta_t$ if any). 
If we include $T_2$ in the simulation times, we can get $\mathsf{Num}_{T_2}$ for the simulated sample paths. 

\section{Generic Simulation Scripting Framework}\label{GenSimFW}

In \cite{PDML}, we introduced a generic simulation scripting framework (GenSimFW) 
to simulate stochastic processes and compute functionals 
based on observed values of those simulated stochastic processes, starting 
from a textual description in close-to-mathematical formulation. This script framework 
can use several backends, currently TensorFlow in computational graph mode, JAX, and NumPy. 
TensorFlow computational graphs generated by the TensorFlow backend are amenable to 
TensorFlow computational graph operations such as \verb|tf.gradients|
which allow one to augment the computational graph to one that also computes 
derivatives of specified output values (such as the payoff) with respect
to any other node of the computational graph including parameters or inputs.  
\cite{PDML} used the scripting framework to introduce a new methodology for parametric 
pricing called parametric differential machine learning which learns conditional expectations given samples
of the output random variate and the conditioning variables and the 
path-wise derivative of the output random variate with respect to the conditioning variables. 

We will explain the features of the input script needed to perform the modeling 
and calibration tests in this paper by discussing the example
scripts in Figures \ref{cheyettescript1} and \ref{cheyettescript2}. 
In Figure \ref{cheyettescript1}, on the top, there are function definitions for the rest of the script. 
(The script framework predefines a set of functions
that is common to the different backends - such as \verb|exp|, 
\verb|oneslike|, \verb|zeroslike|, \verb|sqrt|, and \verb|positivepart| as 
used in the script, but also others such as \verb|einsum|.
Other functions that are or can be implemented in the backends 
together with appropriate derivatives can be added to the framework easily.) 

Next, we describe what is simulated and computed at each time-step. 
If the line starts with \verb|d_|, it describes an increment.
In the right-hand side, \verb|d_t| can be used (and will be defined 
to be the current time step size) and any other name 
starting with \verb|d_| on the right-hand side stands for a 
Brownian increment (normal random variate with zero mean
and standard deviation of one, potentially correlated with 
other normal random variates as specified elsewhere 
in the script) which will be generated and values provided 
when the right-hand side will be evaluated. 
Thus, for a description \verb|d_x = rhs|, \verb|x| after the 
step will be \verb|x+rhs| with \verb|x| the value before 
the time step. Here, this is how \verb|ratex|, \verb|ratey|, 
and \verb|ratevariance| will be time-stepped.
The other expressions for \verb|ratevolatility|, \verb|deltafwd|, 
and \verb|volterm| compute these variables  
as functions of (current values\footnote{I.e., if values at the end 
of the time step have already been computed by preceding lines, 
the value at the end of the time steps is used, otherwise, 
the value at the beginning of the time
step as computed in the previous time step (or as given by or 
computed from initial values) is used.} of) other variables, as long as 
the variable defined in the expression 
does not also occur on the right hand side expression. 
Should the variable occur on both sides, it describes a Markovian
update function - but they have not been used in the example scripts. 
 An example description that could be added to the example
script would be \verb|deltafwdmax=max(deltafwdmax,deltafwd_new)| 
which would keep the running maximum of \verb|deltafwd| in 
\verb|deltafwdmax|. Here, if there is a variable name ending 
in \verb|_new| on the right-hand side, it will receive the new 
value of that variable, while without \verb|_new| it refers to the old value. 

Then, the example script in Figure \ref{cheyettescript1} specifies initial values for the 
time-stepped variables, as indicated by the line starting with \verb|init:|.
If initial value for a variable has not been specified, 
it is computed from the simulation description, if possible; otherwise
an error will result. 

Finally, a set of payoffs is defined. In the loop form shown in Figure \ref{cheyettescript1}, 
at least one of the loop variables must be \verb|t|, and specifies the payment time for the payoff. 
The expressions for the values of the loop variable are evaluated 
when parsing the script (as is the string expression for the name of the payoff),
not when the computation described by the parsed script is executed when 
generating the computational graph or when evaluating the 
expressions or the computational graph. 
\verb|nodiscount| means that 
there is no extra discounting. Instead, one could specify 
\verb|discount| {\em discountfactorexpression} where the payoff
would be multiplied with the value of the {\em discountfactorexpression} . 
For instance, one could simulate 
the stochastic discount factor by 
\verb|d_SDF=-(initfwd(t)+ratex)*SDF*d_t| and use \verb|discount SDF| with 
\verb|SDF| being the discount factor expression. Alternatively, 
one could specify \verb|numeraire| {\em numeraireexpression},
in which case the payoff would be divided by the value of the 
numeraireexpression. As an example, one could
simulate the money market account by \verb|d_MMA = (initfwd(t)+ratex)*MMA*d_t|, and then specify 
\verb|numeraire MMA|. Instead of using the loop form, one can specify a single payoff as in 
the examples in \cite{PDML} like:
\begin{lstlisting}
# payoff
maturity: caplet pays positivepart(pf*exp(cx*ratex[fixingtime]+\
                                          cy*ratey[fixingtime])-khat) \
                 nodiscount 
\end{lstlisting}
where a definition for \verb|maturity| and \verb|fixingtime| needs 
to be provided to the scripting framework, and \verb|maturity| 
(or whatever variable or number given before the \verb|:|) will be used as payment time.
In the payoff, square brackets indicate the time at which variables 
will be observed and then used in the computation.

The next script, Figure \ref{cheyettescript2}, also 
demonstrates the specification of correlations 
in the line \verb|d_W*d_Z = rho|. The expression on the right-hand 
side can contain functions, processes, etc.
and in this way can use for settings with local 
or stochastic or mixed local and stochastic correlation.

The script framework also accepts simulations that involve vectors or tensors in the SDEs and also 
as Brownian increments. For that case, dimensions need to be specified, and there are backend
functions (such as \verb|einsum|) that can operate on such vectors or tensors. 

The scripting framework also allows the definition of time structured computations
where updating of specific variables only occurs at certain time points and 
not at every time step. As a possible example, one could keep track of
a maximum or minimum or sum over values observed at a finite set of
given times, or performance for Cliquet like instruments. 

For a more formal introduction into the Generic Scripting framework, see \cite{PDML}.

In \cite{PDML}, we presented a script for caplet pricing for Cheyette model 
with linear benchmark forward rate local volatility (LinBRLV) with uncorrelated CIR SV. 
Compared to the script in \cite{PDML}, we need to make only minor changes to use 
other functional forms for the local volatility term or alternative forms
of SV.  Examples can be found in the scripts that we just discussed to introduce the scripting framework. 
Figure \ref{cheyettescript1} shows caplet pricing for a
Cheyette model with piece-wise linear benchmark forward rate local volatility (PwLinBRLV) 
together with uncorrelated CIR SV.
Figure \ref{cheyettescript2} shows caplet pricing for a Cheyette model 
with local volatility linear in Cheyette factor $x$ (LinXLV) together
 with correlated lognormal SV with quadratic drift (QDLNSV). 
 Note that, in Figures \ref{cheyettescript1} and 
 \ref{cheyettescript2}, we price caplets 
 with different strikes simultaneously whereas the scripts in 
 \cite{PDML} show caplet pricing for one strike at a time.   
\begin{figure}[h]  
\begin{lstlisting}
# function definition
g(x) = (1/mr)*(oneslike(x)-exp(-mr*x)) 
h(x) = exp(-mr*x)
sigmaFun1(x) = a1*oneslike(x) if x < K1*oneslike(x) else zeroslike(x)
sigmaFun2(x) = a1 + ((a2 - a1)/(K2 - K1))*(x - K1*oneslike(x)) \
	if x >= K1*oneslike(x) and x < K2*oneslike(x) else zeroslike(x)
sigmaFun3(x) = a2 + ((a3 - a2)/(K3 - K2))*(x - K2*oneslike(x)) \
       if x >= K2*oneslike(x) and x < K3*oneslike(x) else zeroslike(x)
sigmaFun4(x) = a3*oneslike(x) if x >= K3*oneslike(x) else zeroslike(x)

#system
d_ratex = (ratey-mr*ratex-g(measT-t)*ratevariance*volterm*volterm)*d_t+\
          ratevolatility*volterm*d_W 
d_ratey = (ratevariance*volterm*volterm-2.0*mr*ratey)*d_t
d_ratevariance = theta*(oneslike(ratevariance) - positivepart(ratevariance))*d_t + volofvar*ratevolatility*d_Z
ratevolatility = sqrt(positivepart(ratevariance))
deltafwd = initfwd(t + delta) + h(delta)*(ratex + g(delta)*ratey)
volterm = sigmaFun1(deltafwd) + sigmaFun2(deltafwd) + sigmaFun3(deltafwd) + sigmaFun4(deltafwd)

#initial values
init: ratex = zeros([batchsize])
init: ratey = zeros([batchsize])
init: ratevariance = ones([batchsize])

#payoffs
for (t,k) in ([maturity]*len(strikes),strikes): "calloption_strike_\%f"\%k \
	pays (positivepart(poa*exp(g(delta)*ratex[t] + \
        0.5*g(delta)*g(delta)*ratey[t]) - 1 - k*delta)) nodiscount
                                          
\end{lstlisting}
\caption{Script for Caplet Pricing for Cheyette Model with uncorrelated CIR SV 
(Euler for Cheyette, Euler full truncation for CIR),
with piece-wise linear benchmark forward rate local volatility (PwLinBRLV). \label{cheyettescript1}} 
\end{figure}

\begin{figure}[h]  
\begin{lstlisting}
#function def 
g(x) = (1/mr)*(oneslike(x)-exp (-mr*x))

#system 
d_ratex = (ratey - mr*ratex - g(measT-t)*volterm*volterm*sigma*sigma)*d_t + volterm*sigma*d_W
d_ratey = (volterm*volterm*sigma*sigma - 2.0*mr*ratey)*d_t
d_sigma = ((kappa1 + kappa2*sigma)*(1.0 - sigma) - g(measT-t)*volterm*beta*sigma*sigma)*d_t + beta*sigma*d_W + eps*sigma*d_Z
volterm = a + b*ratex	

#correlations-
d_W*d_Z = rho

#initial values 
init: ratex  = zeros([batchsize])
init: ratey  = zeros([batchsize])
init: sigma = ones([batchsize])

#payoffs
for (t,k) in ([maturity]*len(strikes),strikes): "calloption_strike_\%f"\%k \
	pays (positivepart(poa*exp(g(delta)*ratex[t] + \
       0.5*g(delta)*g(delta)*ratey[t]) - 1 - k*delta)) nodiscount

\end{lstlisting}
\caption{Script for Caplet Pricing for Cheyette Model with Lognormal SV with quadratic drift 
(Euler for both Cheyette and QDLNSV) with 
local volatility linear in the Cheyette factor $x$ (LinXLV).  \label{cheyettescript2}} 
\end{figure}

Given the appropriate scripts and set-up information, 
GenSimFW provides a variety of features for a variety of 
backends. All start with parsing and preprocessing of the 
script and preparing for interpretation or 
code generation. There is a direct execution mode, in which 
the parsed script is used to execute TensorFlow
computational graph building mode functions to create 
a computational graph that can be further 
worked with (such as adding computation of appropriate derivatives) and then run - 
as in TensorFlow version 1 or the version 1 compatibility layer of TensorFlow version 2, 
or is used to execute appropriate NumPy or JAX functions. 
There is also a code generation mode, in which the parsed script is used 
to generate a Python module containing TensorFlow, NumPy, 
or JAX calls; which then would create a computational graph 
(for the TensorFlow backend) or execute the computation (for NumPy and JAX).
The basic computation supported currently is simulation and 
sample generation in batches (including samples of path-wise derivatives 
generated through augmented computational graphs in TensorFlow 
or appropriately code generated) either with 
fixed parameters or with varying parameters for different 
trajectories in the batch. These samples can 
either be directly used for training in DML or PDML type 
approaches or can be used to compute MC estimates for (conditional) expectations 
and potentially their sensitivities directly by MC averaging. 
These simulations and MC estimates can also use antithetic 
variates, control variate approaches, and similar methodologies.  
There are extensions for calibration, XVA, and 
Bermudan option exercise strategies and 
option pricing which we will describe elsewhere.

\section{Code generated Monte-Carlo pricers and their use within Calibration}\label{MCPCG}


Using GenSimFW as just described, one can execute (or code generate 
and execute) standard Monte-Carlo pricers that generate MC estimates of prices 
given parameter values. One could now run global or local optimizers 
that do not need derivatives (such as variants of differential 
evolution, for instance, ICDE, but there are other global optimization 
packages available in open source, for instance DFOLS by NAG) to find 
optimized parameter sets through minimization of appropriate objective functions. 

However, MC estimates are currently rarely used for calibration 
out of a number of concerns, some related to the inherent complexity while 
others are related directly to the use of Monte Carlo approaches.  
For one, calibration can be provided by some monolithic function or program, 
or it could be provided by some set of components that are appropriately 
connected and steered. There is also a question about how configurable 
the monolithic function or components are. Monolithic designs will 
force more one-off work and even if configurable, all things need to be 
configured in a monolithic manner.  Component-based designs might have to  
be one-off or partially configurable, but the component design allows
one to combine these components more freely.   Given the complexity
 (global/local optimizer, objective function, simulation, 
post-processing), monolithic designs and components will be very complex 
yet need to fit together so that work cannot be easily distributed among 
a team. Monolithic designs also typically generate their random numbers 
internally and make it harder to split generation and consumption 
of random numbers, while in a component design, random number 
generation can be its own component. 
Therefore, one also encounters a more flexible, component based 
design, where components are either one-off or configurable. There has to 
be a (configurable) layer where these components are combined 
appropriately for the pricing and calibration tasks at hand. That control component 
could be a scripting layer embedded in one of the implementation 
languages (C, C++, etc.) using some scripting functionality; or it could be 
a Python interface (or another appropriate scripting language) to the 
components including control component(s). Such a design often also becomes 
complex and is hard to design, develop, and support; 
even though typically less complex as in a monolithic design.  

Developing, managing, supporting, and using one-off implementations for 
different models, instruments, and computational architectures (CPUs, GPUs, distributed computing, etc.) 
will result in a complex setup. It will be difficult and challenging to make sure that 
implementations for different models, instruments, and computational 
architectures will be consistent.  
It will be difficult to train and support developers and 
users to develop and use implementations in such a setting. 

Scriptable and configurable components and set-ups introduce varying 
amounts of overhead in C++ and other implementation languages. If components
can be scripted and configured separately, care has to be taken so 
that scripting and configuration are consistent and that the developers
and users do not have to learn and use different scripting and 
configuration languages for each component. If each component embeds 
a separate scripting layer or interface, it can become challenging 
to coordinate all the different scripting layers and interfaces. 
If each component has a separate Python interface, one now needs to 
build the entire system from Python, with separate knowledge 
of each Python interface, and there is a question how the different 
components can work on shared data or how the data is moved 
between the different components. 

If some or all of the components are written in Python, there is the 
concern that pure Python implementations could be too slow or not 
performant enough, while using standard or specialized libraries like 
NumPy, JAX, TensorFlow, PyTorch, and others requires developers
and users to become familiar enough with the different libraries to
 use them, and in particular to use them efficiently enough for 
purpose.  Similarly, if specialized components are written in C++, 
C, C\#, FORTRAN because of speed or interface concerns, at least 
developers will need to be proficient in these various programming 
languages and development styles, while at the same time 
ensuring that the components will work well together as part of the 
larger system. The challenges become even more substantial if different 
operating systems need to be supported and if performant implementations 
on different target computing architectures will require 
different libraries or set-ups (CUDA for NVIDIA GPU, OpenCL for other GPUs, 
OpenMP for shared-memory distributed computing, MPI for distributed computing, etc.).   

These and other considerations often force the use of a small number 
of paths (1000-5000 in CVA applications) for MC implementations 
because the architecture cannot support more or the architecture 
can only provide these few paths in the given computational budget.
However, MC estimates using such small number of paths are often 
unreliable and exhibit large variance, and often require specialized,
one-off methodologies to address or mitigate reliability and 
variance concerns (for instance, using control variates that are strongly 
tailored to model and instrument, with tests of quality, reliance, 
and accuracy of varying frequency outside of production use).

In this paper, we use code generated GenSimFW parametric MC pricers 
together with other components for some calibrations, and we 
have used TensorFlow computational graphs generated by direct execution 
for some tests. Using GenSimFW in this way 
addresses the above concerns:
\begin{itemize}
\item The GenSimFW based implementations are component based.
\item Everything is steered and configured in one place through 
Python and GenSimFW configuration and set-up. Users in general
do not need to be proficient in the underlying libraries or 
handle different aspects beyond configuration and scripts.
\item Random numbers (Pseudo- or Quasi- or Pseudorandomized Quasi-Random numbers) 
can be generated by the component internally, by an external component, 
or precomputed in any fashion and provided to the component from outside. 
\item For global or local optimization, we can use outside optimizers 
implemented in or interfaced to in Python, such as the SciPy 
optimization suite or openly accessible ones like NAG's DFOLS, 
or use internally developed ones 
(we developed and are using ICDE as a differential evolution variant in our case).
\item Using GenSimFW allows one to separate concerns, large parts of 
the complexity are addressed by the framework and as such 
do not need to be addressed by users of the framework.
\item GenSimFW translates the script which is written in a 
common API that is supported by all its backends (and could be supported 
by others libraries as well) and is then translated into 
appropriate TensorFlow 1\footnote{Or version 1 compatibility part of TensorFlow 2},
NumPy, or JAX function calls so that users do not need to know how 
to implement in TensorFlow, NumPy, or JAX. GenSimFW is extensible 
and additional functions can be provided within or outside the script and/or framework. 
\item The current backends TensorFlow, NumPy, and JAX run on many 
current computational architectures (CPU, GPU, TPU, Intel, PowerPC, etc.)
and different operating systems and are highly optimized and vectorized. 
Most time is spent in computational kernels and data handling, not in 
scripting. There are optimized versions for particular settings 
and architectures that still provide the same general API. There are also
optimizing variants such as Numba or similar. The NumPy interface is 
based on data structures that are common and shared between Python and 
C/C++ versions so that data can be efficiently communicated, transferred, 
and used in different languages. Many libraries provide a partial or complete 
NumPy implementation. Typically, the libraries rely on BLAS, XLA, 
and similar library interfaces, and more efficient and specialized 
versions can be used that implement the same interface. 
\item Different backends provide additional features. For instance, 
TensorFlow 1 provides optimizations and compilation for computational 
graphs\footnote{Native TensorFlow 2 provides optimization and compilation 
for Python scripts that use TensorFlow functions, through 
tracing.}, TensorFlow 1 has algorithmic differentiation on the 
level of computational graphs\footnote{TensorFlow 2 implements algorithmic 
differentiation through tracing and tapes, which incurs 
additional overhead and work and makes adding higher 
order derivatives much less convenient compared to TensorFlow 1.}
 and JAX has JIT compilation to XLA, vectorization, and algorithmic 
differentiation\footnote{All implemented through tracing and optimizing over 
a different, internal, representation JaxPR, which can 
not be generated easily from the outside or executed easily by itself.}. 
In this way, we can take advantage of additional features of 
the backends in a generic way. Different backends also have ways to run 
different precisions and bitnesses, allowing lower precision 
results to be computed substantially faster.
\item Using GenSimFW in the direct execution mode, we have overhead 
in the generation of the TensorFlow computational graph and in 
the execution of the NumPy or JAX computation (since direct execution 
combines both parsing and interpretation of the script with
execution of the actual computation or graph generation), but not in 
the TensorFlow computations based on the generated computational graph. 
\item Tracing through the direct execution mode functions will result 
in complicated traces that do not reflect a good trace of the actual computation. 
GenSimFW could directly generate computational graphs, ASTs, or 
tapes for libraries that support algorithmic differentiation or 
Just-In-Time compilation (just as it generates computational graphs 
in and for TensorFlow1) but unfortunately XLA, JaxPR etc. cannot be easily generated 
and worked with from outside JAX or TensorFlow. 
\item However, using GenSimFW in code generation mode, using the generated modules 
results in no or minimal overhead, since the generated
code are specialized as much as possible. 
The generated modules are straight-line code and are quite efficient, 
leading to simulation and pricing that is sufficiently fast for many 
purposes, with sizeable batches of paths computed at the same time, 
depending on model and instrument complexity ranging from 50,000 to more than 
a million or more paths. 
\item Being Python with well-known libraries, the generated code can be 
inspected and debugged easily on its own, with standard Python editors and debuggers. 
\item Computational graph generation in TensorFlow 1 is more efficient 
through the code generated Python-TensorFlow scripts than through
the direct execution, but not dramatically so.
Appropriately code generated Python JAX scripts can be traced, 
JITed, JAX vectorized, and JAX differentiated, trading  
time and memory needs for tracing, JIT, and algorithmic 
differentiation for much more efficient execution and the computation of 
appropriate derivatives, subject to limitations imposed by JAX and its implementations. 
\end{itemize}

\begin{figure}[h]
    \centering
    \fbox{
    \begin{minipage}{0.95\textwidth}
    \begin{itemize}
    \item Input: Script and set-up information
    \item Preprocessing script, classifying \& splitting, parsing appropriate parts, 
    analyze, rewrite 
    \item Direct execution mode: 
    \begin{itemize}
    \item TensorFlow 1 and TensorFlow 2 Version 1 compatibility layer 
    \begin{itemize}
    \item Generate TF1 computational graphs (CGs)
    \item User can post-process and/or augment CGs with TF (e.g. algorithmic differentiation)
    \item User runs CGs in TF, extracts, and post-processes results 
    \end{itemize}
    \item NumPy or JAX 
    \begin{itemize}
        \item Call appropriate NumPy or JAX functions to perform computations 
        \item User extracts and post-processes results 
    \end{itemize}
    \end{itemize}
    \item Code generation mode:
    \begin{itemize}
        \item TensorFlow 1 and TensorFlow 2 Version 1 compatibility layer 
        \begin{itemize}
        \item Generate specialized code that will generate TF1 CGs
        \end{itemize}
        \item NumPy or JAX 
        \begin{itemize}
            \item Generate specialized code that will call appropriate NumPy or JAX functions to perform computations 
            \item Generated code is straight line and easier to trace and debug (so user can use JAX features or 
            debugger)
        \end{itemize}
        \end{itemize}
    \end{itemize}
    \end{minipage}}
\caption{GenSimFW modes}
\label{gensimfwmodes}
\end{figure}

\begin{figure}[h!]
    \begin{center}
    \begin{tikzpicture}

    \node[draw] (mdata) at (0,0) {\begin{tabular}{l} Market Data \\ \hspace*{5mm} Curves etc. \\ 
                                    \hspace*{5mm} Prices of Calibration Instruments \end{tabular}};
    \node[draw] (parampricer) at (0,4) {\begin{tabular}{l} Parametric Pricer \\ \hspace*{5mm} Generated code, \\ 
                                        \hspace*{5mm} Generated CG, \\ \hspace*{5mm} or PDML \end{tabular}};
    \node[draw] (objfun) at (4,2) {\begin{tabular}{l} Objective Function \\ \hspace*{5mm} Sum of squares of residual \end{tabular}}; 
    \node[draw] (globlocopt) at (8,4) {\begin{tabular}{l} Global or local optimizer \\ \hspace*{5mm} E.g. SciPy, ICDE, \\
                                        \hspace*{5mm}  NAG's DFOLS, etc. \end{tabular}}; 
    \node[draw] (optparams) at (4,6) {Optimized parameters};
    
    \draw[->] (mdata) to (objfun);
    \draw[->] (parampricer) to (objfun);
    \draw[->] (objfun) to (globlocopt); 
    \draw[->] (globlocopt) to (optparams);
    \draw[->] (optparams) to (parampricer);
        
\end{tikzpicture}
\end{center}
\caption{Calibration with Parametric Pricers \label{calibrationfig}}
\end{figure}
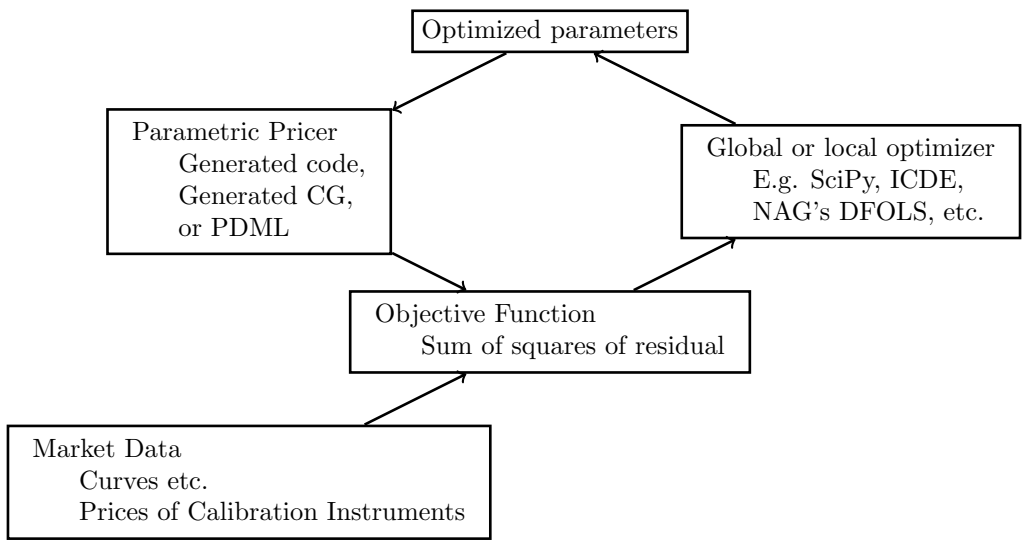

We thus obtain parametric pricers and calibration set-ups with a much 
lower barrier to entry and a much simplified implementations of models
and instruments, allowing us to simulate and price a large variety of 
models, and calibrate a large variety of models sufficiently 
efficiently in this way (in a minute or less in many of our tests).

\section{Learning Parametric Pricers through Parametric Differential Machine Learning for Pricing and Calibration}\label{PDMLIntro}


Often, one has some stochastic model (such as a system of SDEs, 
ODEs, and other Markovian updates) with one or several model parameters,
and one is interested in the efficient estimation of expectations 
and conditional expectations (in quantitative finance, these
are current prices or future (conditional) prices). For 
situations such as counterparty 
or market risk, one often needs expectations of payoffs 
conditional on the Markovian state at some intermediate time. 
Savine and Huge demonstrated the potential of a Differential 
Machine Learning technique, where a neural network 
is trained on samples of the payoff, the Markovian state to 
be conditioned on, and the path-wise derivatives of
the payoff with respect to the Markovian state. 
For many stochastic processes and settings (corresponding
to parabolic equations satisfying certain conditions), 
the dependency on state becomes smooth after 
infinitesimal small time. However, one is often 
interested in the behavior of the conditional expectation 
in dependence of the model parameters or parameters 
involved in the payoff, not only conditional 
on the Markovian state. Therefore, we introduced 
Parametric Differential Machine Learning (PDML)
in \cite{PDML} where we vary model and/or 
contract parameters either without varying 
Markovian state or together with varying Markovian state.

More formally, parametric differential machine learning 
(PDML) approximates the conditional expectation $E[Y | X]$ 
by training a deep neural network $N(X;\Theta)$  
with parameters $\Theta$ (typically weights and biases)
on $N_{S}$ joint samples $\{ \left(X^i,Y^i, DY^i \right), i = 1, ..., N_S \}$
with $X^i = \left( X^{i}_{j}, j = 1, ..., N_X \right)$ 
samples of the conditioning variables (Markovian state, model parameters, 
and/or contract parameters) each of size $N_X$,  
$Y^i = \left( Y^{i}_{j}, j = 1, ..., N_Y \right)$ corresponding samples
of the output variable(s) -- typically payoffs -- each of size $N_Y$, and 
$DY^i$ denotes the collection of path-wise differentials $\frac{\partial Y_{J}^{i}}{\partial X_{I}^{i}}$ 
where $I$ and $J$ specify which partial differentials of output variables with respect to which conditioning variables
are used in the training. 
The loss function in PDML is
\begin{equation}\label{PDMLlosseqn}
\Theta_N^{*,PDML} =  \underset{\Theta_N}{\arg\min} E\left[\left| Y - N(X;\Theta_N) \right|^2 
+ \sum_{i,j} \lambda_{ij} \left| DY_{ij} - \frac{\partial N(X;\Theta_N)_{I(i)}} {\partial X_{J(j)}} \right|^2 \right]. 
\end{equation}
We denote $DY_{ij}=\frac{\partial Y_{I(i)}}{\partial X_{J(j)}}$ 
where mappings $I$ and $J$ specify which partial derivatives are used in the loss function.

In \cite{PDML}, we showed how PDML can be trained to learn the 
price of caplets as a function of contract and/or model parameters. 
We can use uniform sampling distributions to sample model 
or contract parameters for the parametric simulation 
to generate samples with those conditioning variables. 
However, if the magnitude of $Y$ varies substantially for 
different parameter regions in the parameter set, 
training on uniformly sampled parameters might lead 
to worse fit in the parameter regions where the
 magnitude of $Y$ is smaller. In such cases,
one can use adaptive sampling to improve fit 
in such regions, as demonstrated in \cite{PDML}.
In \cite{PDML}, we observed that using differentials 
in the training leads to faster convergence of price predictions
and risk sensitivities predictions and 
to increased accuracy of prices and sensitivities.

Once trained, the deep neural networks can be used 
as fast parametric surrogates and pricers. The parametric pricer
so obtained can be used inside a global optimizer such as 
ICDE to find optimized parameter sets for a variety
of calibration objective functions. Such a calibration 
process uses random seeds at several points, when 
generating parameters for the sampling, simulating 
the stochastic processes, initializing weights and biases
for the training of the DNN, and when ICDE generates new 
populations by random sampling and cross-over. 
Different seeds will in general lead to 
different optimized parameter sets that lead to different 
accuracies. Instead of trying to minimize randomness, 
we can use randomness and replication 
to obtain more robust calibration processes. 
We test the accuracy of the parametric pricers
by some Monte-Carlo estimates on the last 
optimized parameters. Based on these indicators of 
accuracy, we studied two approaches, a 
`best seed' approach and an ensemble approach,
to obtain a more robust calibration based 
on several random replications of training 
a surrogate and calibration by optimizing 
over that surrogate. We used these calibration
approaches to calibrate models with 
constant parameters against single maturity 
data and to calibrate models with piece-wise 
constant parameters against data
for several maturities, with a bootstrapping approach 
where robustified calibration as just discussed is 
used to determine parameters up to the 
first maturity and then between maturities.
We demonstrated these calibration approaches in \cite{PDML} on the example of 
linear benchmark forward rate local volatility with uncorrelated CIR SV.

In this paper, we will apply these approaches to calibrate some 
One Factor Cheyette models with various functional forms for the local volatility 
combined with various stochastic volatility or variance SDEs.

\section{Calibration using code generated MC pricers vs. calibration with PDML}\label{whichway}

PDML calibration approaches require implementations where derivatives are well-behaved, ML approaches require
adaptations so that training of surrogates is fast, efficient, and converges well and where convergence
and accuracy can be easily checked. In addition, these approaches sometimes only work really well if there are 
not too many parameters or where dimensionality reduction techniques can be applied such as differential PCA. 
Also, for instance, if the location of interpolation points are parameters, derivatives with respect to the 
interpolation point locations typically cannot be computed by the standard algorithmic differentiation
of computations. Computations for different locations or numbers of interpolation points often 
require different implementations. There are also cases in which the results do not smoothly depend on at 
least some of the parameters, either locally or globally. 
Such situations pose particular challenges for PDML and similar approaches,
complicating the sampling but also the accurate computation and use of derivatives. 
When PDML methods cannot be applied as easily (or the user or developer does not want to go through
the trouble and effort to implement and to test such methods), calibration around code generated MC
pricers can be an efficient and useful alternative. Its use provides great flexibility 
and convenience and allows straightforward calibrations in settings that were previously 
infeasible or too inefficient. 

For instance, for the piece-wise linear benchmark forward rate 
local volatility setting where we optimize over the 
different pieces, PDML approaches can be challenging to apply, 
but calibration around code-generated MC
pricers is straightforward and applicable. Thus, we applied calibration around code-generated MC pricer code
for this and several of the other settings, using derivative-free ICDE variants for the work in this paper.

\section{Numerical Results: Pricing and Calibration}\label{Results}

\begin{figure}[h]
\centering
\includegraphics[width=0.60 \linewidth]{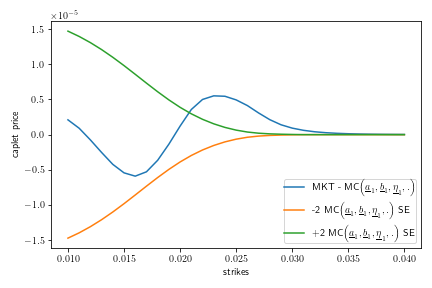}
\caption{1yr Maturity Calibration using Best Seed PDML Calibration Approach for Cheyette Model with Linear Benchmark Forward Rate Local Volatility (LinBRLV) 
and CIR SV. Figure from \cite{PDML}.}
\label{pdmlpaper1yrmaturityexample}
\end{figure}

In our previous paper \cite{PDML}, we demonstrated the calibration 
of caplets with tenor 3M for maturities 
ranging from 1yr to 6yr for Cheyette model with linear benchmark 
forward rate local volatility (LinBRLV) and (uncorrelated) CIR SV. 
In particular, we showed the calibration results for single-seed 
as well as robustified optimizations. 
We also considered model calibrations to single maturities 
as well as piece-wise calibrations to several maturities.  
We observed that except for 1yr maturity, the price difference 
between market prices from the CapFloor volatility surfaces 
and MC prices with calibrated parameters is well within 2 
standard error bars for most of the strikes and all the strikes near ATM. 
For the 1yr maturity, the Cheyette model with linear benchmark 
forward rate local volatility could not be calibrated well
across all strikes - it was not able to reprice the middle 
strikes (200-300bps) at the same time as repricing the other strikes and the tails, 
as shown in Figure \ref{pdmlpaper1yrmaturityexample}.
Here, MKT denotes the market price of the caplet for the 
corresponding strike and MKT - MC  shows how well 
the optimized parameter set with enough samples of MC 
(approximate ground truth) approximates the MKT prices. 
Here, the optimized parameter set is obtained by best seed 
PDML calibration approach proposed in our previous paper \cite{PDML}.
In Figure \ref{pdmlpaper1yrmaturityexample}, we can see that approximate 
ground truth values obtained using optimized parameter 
set did not reprice target prices (MKT) well in some ranges which 
could indicate some limitations and properties of the chosen model setting.

\begin{figure}
	\centering
	\subfigure[]{\includegraphics[width=0.4\textwidth,height=0.25\textwidth]{figs/pricediffgraph_bestseed_20201006_Modelerror.png}}
	\subfigure[]{\includegraphics[width=0.4\textwidth,height=0.25\textwidth]{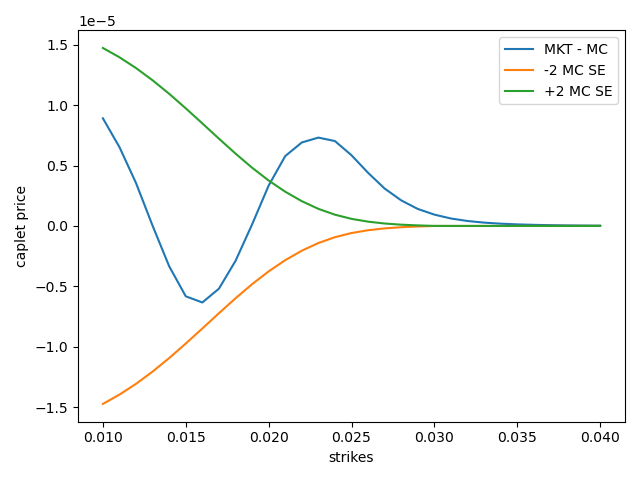}}
	\subfigure[]{\includegraphics[width=0.4\textwidth,height=0.25\textwidth]{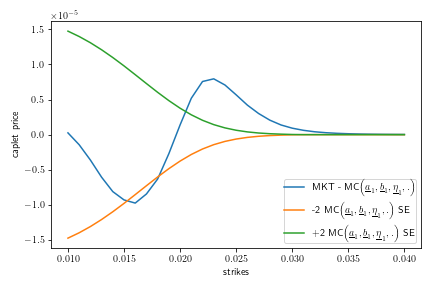}}
	\subfigure[]{\includegraphics[width=0.4\textwidth,height=0.25\textwidth]{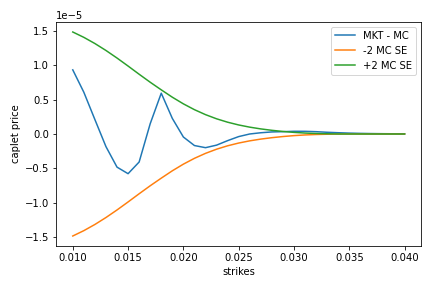}}
	\subfigure[]{\includegraphics[width=0.4\textwidth,height=0.25\textwidth]{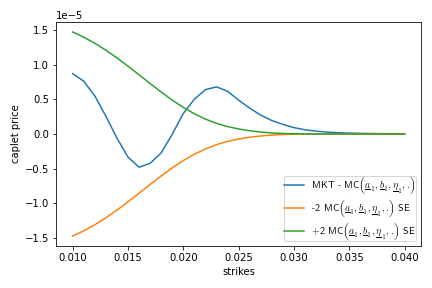}}
	\subfigure[]{\includegraphics[width=0.4\textwidth,height=0.25\textwidth]{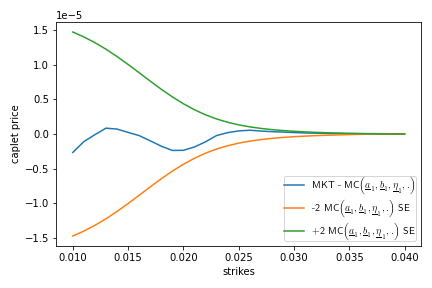}}
	\subfigure[]{\includegraphics[width=0.4\textwidth,height=0.25\textwidth]{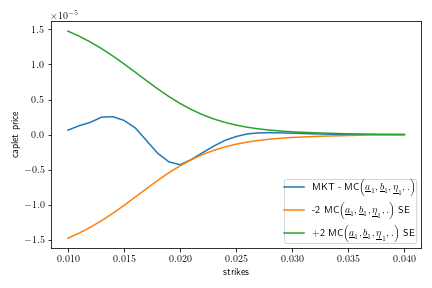}}
	\subfigure[]{\includegraphics[width=0.4\textwidth,height=0.25\textwidth]{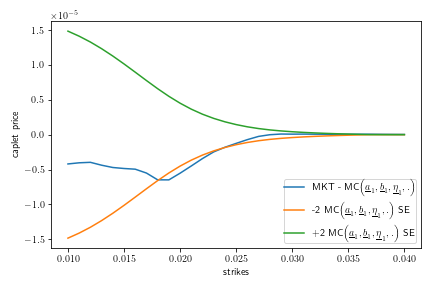}}
	\caption{Calibration results for 1yr maturity caplets under various model settings. 
    (a) LinBRLV + CIR SV (b) LinBRLV (c) LinBRLV + CorCIRSV (d) PwLinBRLV + CIR SV (e) LinSRLV + CIR SV
     (f) LinXLV + QDLNSV (g) LinBRLV + QDLNSV (h) LinSRLV + QDLNSV.}\label{summaryfig}   
\end{figure}

In this paper, we consider Cheyette model with other functional forms for 
the local volatility and with uncorrelated and correlated CIR SV and with correlated  
lognormal SV with quadratic drift (QDLNSV). We investigate whether 
calibration in these different model settings improves the 
repricing of the market volatility slice compared to the
less-than-perfect repricing for 1Y caplets on 3M rates under the 
model setting in \cite{PDML}.  Figure \ref{summaryfig} summarizes the calibration quality 
of a set of such forms considered in this paper.  

\begin{table}[h!]
    \centering
    \begin{tabular}{ |c|c|}
    \hline
    Model Setting & Fixed Parameters \\
    \hline
    LinBRLV + CIRSV &  $\lambda$ = 0.03, $\theta$ = 0.2 \\
    \hline
    LinBRLV & $\lambda$ = 0.03, $\theta$ = 0.2 \\
    \hline
    LinBRLV + CorCIRSV & $\lambda$ = 0.03, $\theta$ = 0.2 \\
    \hline
    PwLinBRLV + CIRSV & $\lambda$ = 0.03, $\theta$ = 0.2, $K_1$ = 0.95ATM, $K_2$ = ATM, $K_3$ = 1.05ATM \\
    \hline
    LinSRLV + CIRSV &  $\lambda$ = 0.03, $\theta$ = 0.2 \\
    \hline
    LinXLV + QDLNSV &  $\lambda$ = 0.025, $\kappa_1$ = 0.25, $\kappa_2$ = 0.25 \\
    \hline
    LinBRLV + QDLNSV & $\lambda$ = 0.025, $\kappa_1$ = 0.25, $\kappa_2$ = 0.25 \\
    \hline
    LinSRLV + QDLNSV & $\lambda$ = 0.025, $\kappa_1$ = 0.25, $\kappa_2$ = 0.25 \\
    \hline
    \end{tabular}
    \caption{Fixed parameters for each model setting shown in Figure \ref{summaryfig}.}
    \label{table:1fixed}
\end{table}

\begin{table}[h!]
    \centering
    \begin{tabular}{ |c|c|}
    \hline
    Model Setting & Parameter Bounds \\
    \hline
    LinBRLV + CIRSV &   $a :[0.0001, 0.015]$, $b:[-0.5, 0.5]$, $\eta$\tablefootnote{\label{fellercondnote} The parameter bounds for the volatility of 
    variance are chosen such that the Feller condition is satisfied.}:$[0.1, 0.63]$ \\
    \hline
    LinBRLV & $a:[0.0001, 0.015]$, $b:[-0.5, 0.5]$ \\
    \hline
    LinBRLV + CorCIRSV & $a:[-0.1, 0.1]$, $b:[-0.1, 0.1]$, $\eta$\textsuperscript{\getrefnumber{fellercondnote}}:$[0.1, 0.63]$, $\rho:[-0.9, 0.9]$ \\
    \hline
    PwLinBRLV + CIRSV &  $a_1:[10^{-6}, 0.5]$, $a_2:[10^{-6}, 0.5]$, $a_3:[10^{-6}, 0.5]$,  $\eta$\textsuperscript{\getrefnumber{fellercondnote}}:$[0.1, 0.63]$\\
    \hline
    LinSRLV + CIRSV &  $a:[0.0001, 0.015]$, $b:[-0.5, 0.5]$, $\eta$\textsuperscript{\getrefnumber{fellercondnote}}:$[0.1, 0.63]$ \\
    \hline
    LinXLV + QDLNSV &  $a:[-0.1, 0.1]$, $b:[-0.1, 0.1]$, $\beta:[-0.1, 0.1]$, $\epsilon:[0.1, 1.0]$ \\
    \hline
    LinBRLV + QDLNSV & $a:[-0.1, 0.1]$, $b:[-0.1, 0.1]$, $\beta:[-0.1, 0.1]$, $\epsilon:[0.1, 1.0]$ \\
    \hline
    LinSRLV + QDLNSV &  $a:[-0.1, 0.1]$, $b:[-0.1, 0.1]$, $\beta:[-0.1, 0.1]$, $\epsilon:[0.1, 1.0]$ \\
    \hline
    \end{tabular}
    \caption{Parameter bounds used in calibration for each model setting shown in Figure \ref{summaryfig}.}
    \label{table:1parambnds}
\end{table}

\begin{table}[h!]
\centering
\begin{tabular}{ |c|c|c|}
\hline
Model Setting & Calibrated Parameters & Good Fit? \\
\hline
LinBRLV + CIRSV & $a$ = 0.00832, $b$ = -0.19208, $\eta$ = 0.56724 & \tikzxmark \\
\hline
LinBRLV & $a$ = 0.00762, $b$ = -0.15945 & \tikzxmark \\
\hline
LinBRLV + CorCIRSV & $a$ = 0.00679, $b$ = -0.09999, $\rho$ = -0.46437, $\eta$ = 0.31777 & \tikzxmark \\
\hline
PwLinBRLV + CIRSV & $a_1$ = 0.00615, $a_2$ = 0.00361, $a_3$ = 0.00511, $\eta$ = 0.62762 & \tikzcmark\\
\hline
LinSRLV + CIRSV & $a$ = 0.00766, $b$ = -0.15588, $\eta$ = 0.34177 & \tikzxmark \\
\hline
LinXLV + QDLNSV & $a$ = -0.00522, $b$ = 0.08982, $\beta$ = 0.09999, $\epsilon$ = 0.55189 & \tikzcmark\\
\hline
LinBRLV + QDLNSV & $a$ = -0.00665, $b$ = 0.09999, $\beta$ = 0.09999, $\epsilon$ = 0.60884 & \tikzcmark\\
\hline
LinSRLV + QDLNSV & $a$ = -0.00674, $b$ = 0.09999, $\beta$ = 0.09999, $\epsilon$ = 0.62731 & \tikzxmark\\
\hline
\end{tabular}
\caption{Calibrated parameters for each model setting shown in Figure \ref{summaryfig}.}
\label{table:1}
\end{table}

From the summary figure, we see that the model settings in panels 
(d), (f), and (g) lead to calibrations that reprice volatility slice market data 
within two Monte Carlo Standard Error estimates, while the 
other settings do not. We will now discuss particular calibrations in turn. 

\begin{figure}[h]
\centering
\includegraphics[width=0.60 \linewidth]{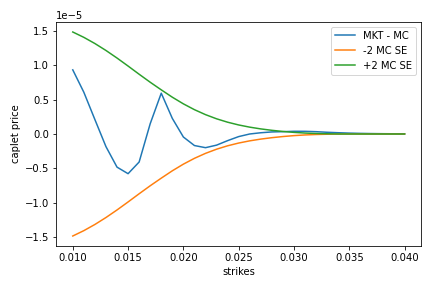}
\caption{1yr Maturity Calibration using MC Calibration for Cheyette Model with Piece-wise Linear Benchmark Forward Rate Local Volatility form (PwLinBRLV) 
and uncorrelated CIR SV.}
\label{1yrbfpcl_MC}
\end{figure}

Firstly, as a straightforward extension of the Cheyette model setting as in \cite{PDML}, 
we consider a Cheyette model with piece-wise linear benchmark 
forward rate local volatility (PwLinBRLV) together with an uncorrelated CIR SV. 
We use an MC calibration approach with code generation for this 
model setting. We chose a set of 3 strikes to form a piece-wise linear function
and these strikes were chosen based on the ATM strike, in particular, we chose
$K1$ = 0.95$\times$ATM, $K2$ = ATM, and $K3$ = 1.05 $\times$ATM. 
In Figure \ref{1yrbfpcl_MC} we show the calibration result 
of 1yr maturity for this new model setting.  
In this figure, MC denotes the MC with enough samples computed 
using the optimized parameter set obtained by MC calibration approach. 
We can clearly see that the price difference between MKT and MC is 
well within 2 standard error bars for all the strikes. This model setting
improves the repricing and calibration quality over the previous model 
setting. However, choosing the appropriate set of strikes to use to define
the piece-wise linear function can be a challenge and 
one has to choose them by trial and error.
It would be good to have a model setting in which one does not need 
to choose such details ad-hoc or such details are picked automatically 
while still leading to consistent results. 

\begin{figure}[h]
\centering
\includegraphics[width=0.60 \linewidth]{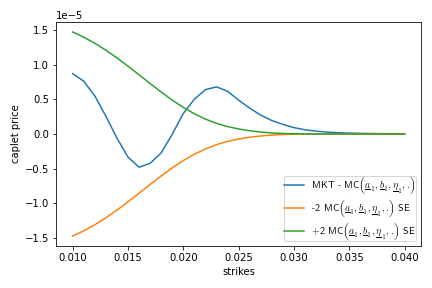}
\caption{1yr Maturity Calibration using Best Seed PDML Calibration Approach for Cheyette Model with Linear Short Rate Local Volatility form (LinSRLV) 
and uncorrelated CIR SV.}
\label{1yrshortrate_pdml}
\end{figure}

 Next, we consider a linear short rate local volatility functional f
 orm (LinSRLV) together with uncorrelated CIR SV. 
 The calibration result is shown in Figure \ref{1yrshortrate_pdml}. 
 We can see that this linear short rate volatility functional form 
 with uncorrelated CIR SV does not replicate the market prices in the middle of 
 the plotted strikes well, similar to the results we obtained in \cite{PDML}. 

\begin{figure}[h]
\centering
\includegraphics[width=0.60 \linewidth]{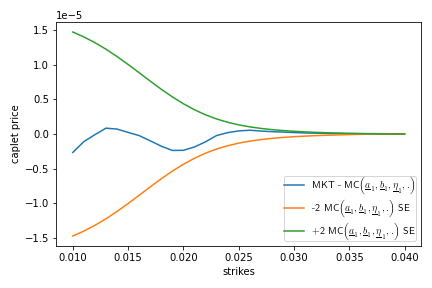}
\caption{1yr Maturity Calibration using Best Seed PDML Calibration Approach for Cheyette Model with Local Volatility Linear in Cheyette Factor (LinXLV) 
and Correlated Lognormal SV with Quadratic Drift (QDLNSV).}
\label{1yrlocalvol_LNSV_pdml}
\end{figure}

Then, we consider the linear Cheyette factor local volatility functional 
form (LinXLV) with a correlated lognormal SV with quadratic drift (QDLNSV). 
This form of lognormal SV was introduced in \cite{LNSVSeppetal_2023}. 
The calibration result is shown in Figure \ref{1yrlocalvol_LNSV_pdml}. 
We can see that the price difference between MKT and MC is well 
within 2 standard error bars for all the strikes 
and thus fits market prices well even in the region where linear 
benchmark forward rate local volatility and uncorrelated CIR SV
was unable to do so.  

From our numerical investigation so far, we observe that the piece-wise linear 
benchmark forward rate local volatility form (PwLinBRLV) 
with uncorrelated CIR SV  and two forms of local volatility 
(linear Cheyette factor local volatility - LinXLV - or 
linear benchmark rate local volatility - LinBRLV) with 
lognormal SV with quadratic drift (QDLNV) are the only 
forms which priced the 1yr maturity close to the market prices across all the strikes. 

\begin{figure}[h]
\centering
\includegraphics[width=0.60 \linewidth]{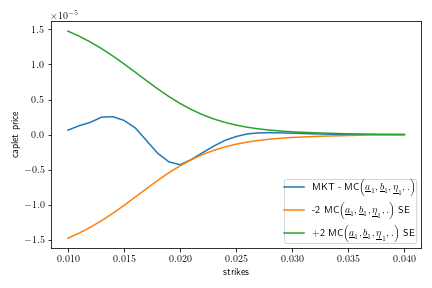}
\caption{1yr Maturity Calibration using Best Seed PDML Calibration Approach for Cheyette Model with Linear Benchmark Rate Volatility form (LinBRLV) 
and Lognormal SV with Quadratic Drift (QDLNSV).}
\label{1yrbenchmarkvol_LNSV_pdml}
\end{figure}

\begin{figure}[h]
\centering
\includegraphics[width=0.60 \linewidth]{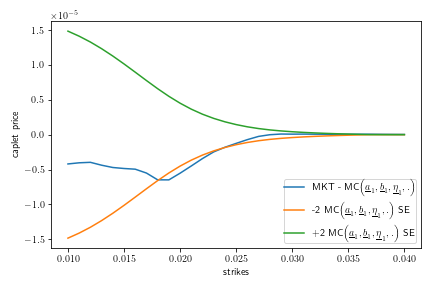}
\caption{1yr Maturity Calibration using Best Seed PDML Calibration Approach for Cheyette Model with linear Short Rate Volatility form (LinSRLV) 
and Lognormal SV with Quadratic Drift (QDLNSV).}
\label{1yrshortratevol_LNSV_pdml}
\end{figure}

Next, we investigate if we can calibrate well using benchmark forward rate 
and short rate volatility functional forms when used with the 
lognormal SV with quadratic drift (QDLNSV). 
Figures \ref{1yrbenchmarkvol_LNSV_pdml} and 
\ref{1yrshortratevol_LNSV_pdml} show the calibration results for 
benchmark rate volatility functional form with QDLNSV and 
short rate volatility functional form with QDLNSV, respectively. 
We can see that in Figure \ref{1yrbenchmarkvol_LNSV_pdml} benchmark 
rate volatility functional form with QDLNSV repriced 
market volatility surface prices well within 2 standard error 
bars, thereby improving accuracy over benchmark rate vol model 
with uncorrelated CIR SV. In Figure \ref{1yrshortratevol_LNSV_pdml},
 we can see that the short rate volatility form with 
QDLNSV priced close to 2 standard error bars in the middle 
of the strikes and better fit when compared to short rate vol 
with uncorrelated CIR SV in Figure \ref{1yrshortrate_pdml}.
 So, this shows for 1yr maturity for caplets, 
lognormal SV with quadratic drift can fit the vol 
smile well and better than CIR SV can.  

\begin{figure}
    \centering
    \subfigure[]{\includegraphics[width=0.49\textwidth]{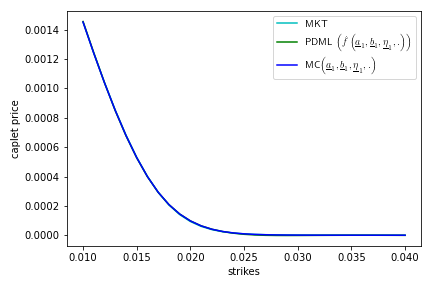}} 
    \subfigure[]{\includegraphics[width=0.49\textwidth]{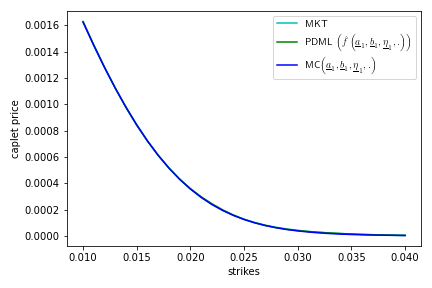}} 
    \subfigure[]{\includegraphics[width=0.49\textwidth]{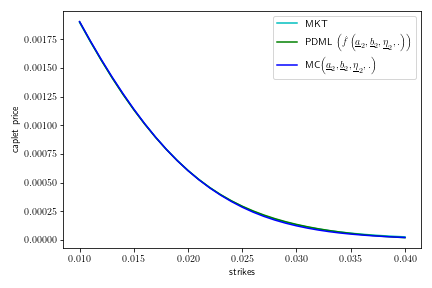}}
    \subfigure[]{\includegraphics[width=0.49\textwidth]{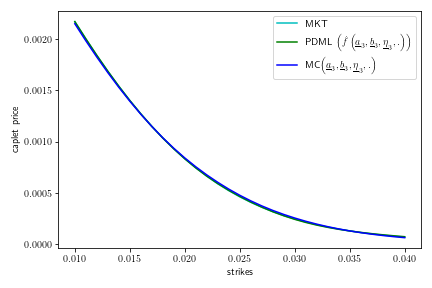}}
    \subfigure[]{\includegraphics[width=0.49\textwidth]{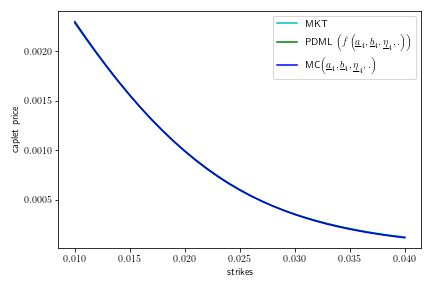}}
    \subfigure[]{\includegraphics[width=0.49\textwidth]{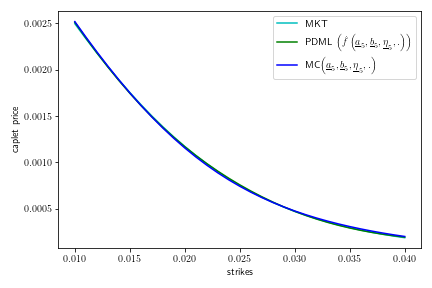}}
    \caption{Calibration with Best Seed PDML Calibration Approach for Cheyette Model with Linear Benchmark Rate Volatility (LinBRLV) 
    and Correlated Lognormal SV with Quadratic Drift (QDLNSV). 
    Plots show Prices from MC (approximate ground truth) at Optimized Parameters Compared to Market Prices for various Maturities. 
    (a) 1yr Maturity (b) 2yr Maturity (c) 3yr Maturity (d) 4yr Maturity (e) 5yr Maturity (f) 6yr Maturity.} 
    \label{pricegraphs_bestseed_benchmarkLNSV}
\end{figure}

\begin{figure}
    \centering
    \subfigure[]{\includegraphics[width=0.49\textwidth]{figs/pricediffgraph_bestseed_20201006_benchmarkvolLNSV_Modelerror.png}} 
    \subfigure[]{\includegraphics[width=0.49\textwidth]{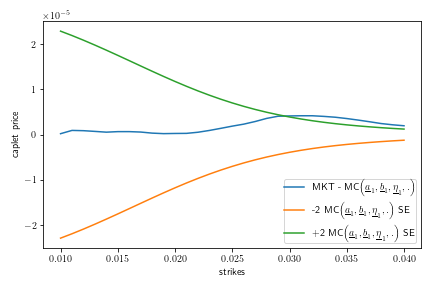}} 
    \subfigure[]{\includegraphics[width=0.49\textwidth]{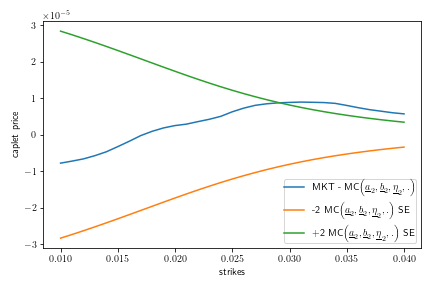}}
    \subfigure[]{\includegraphics[width=0.49\textwidth]{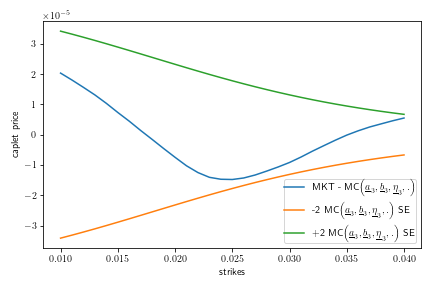}}
    \subfigure[]{\includegraphics[width=0.49\textwidth]{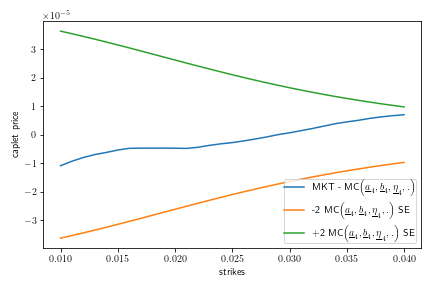}}
    \subfigure[]{\includegraphics[width=0.49\textwidth]{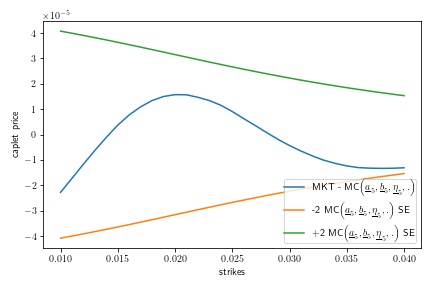}}
    \caption{Calibration with Best Seed PDML Calibration Approach for Cheyette Model with Linear Benchmark Rate Volatility form (LinBRLV) 
    and Correlated Lognormal SV with Quadratic Drift (QDLNSV). 
    Plots show Price differences between MC (approximate ground truth) at Optimized Parameters and Market Prices for various Maturities. 
    (a) 1yr Maturity (b) 2yr Maturity (c) 3yr Maturity (d) 4yr Maturity (e) 5yr Maturity (f) 6yr Maturity.} 
    \label{pricediff_graphs_bestseed_benchmarkvolLNSV}
\end{figure}

Finally, in our previous paper \cite{PDML} we observed that the linear 
benchmark rate volatility functional form  (LinBRLV) with uncorrelated CIR SV 
calibrates well and close to 2 standard error bars for maturities ranging from 2yr to 6yr. 
Here, we investigate whether we can calibrate well for later 
maturities (2yr to 6yr) with linear benchmark rate volatility form (LinBRLV) with QDLNSV,
in addition to the 1yr calibration we already covered. 
We follow the calibration approach outlined in \cite{PDML} for 
calibrating multi-maturity data and settings. 
Figure \ref{pricegraphs_bestseed_benchmarkLNSV} compares the 
price graphs of caplets of maturities ranging from 1yr to 6yr 
from MC with enough samples (treated as approximate ground truth) and market prices.  
Figure \ref{pricediff_graphs_bestseed_benchmarkvolLNSV} shows 
the price differences between MC (approximate model ground truth) 
and market prices for various maturities. 
Both of these figures show linear benchmark rate volatility 
functional form together with correlated QDLNSV calibrated well within 
or close to 2 standard error bars for all the strikes and for all the maturities considered.  

\section{Conclusions}\label{Conclusion}

In this paper we studied whether one-factor Cheyette models with different forms of local 
volatilities and/or stochastic volatilities can calibrate to short maturity caplet smiles
across a wide range of strikes, unlike the one-factor Cheyette model with linear 
benchmark rate local volatility and uncorrelated CIR stochastic variance studied 
in \cite{PDML}. We tested several local volatility forms and stochastic volatility and variance forms. 
We could do so conveniently and efficiently through the generic simulation scripting 
framework and two calibration approaches - one calibrating around parametric 
simulation prices from code generation or computational graph generation as introduced 
in this paper, while the other uses parametric differential machine learning (PDML) as 
introduced in \cite{PDML}. The scripting framework allows the easy and convenient
specification of the models and instruments and the code generation and calibration 
set-ups lead to calibrations that work well with relatively small computational effort 
and often return calibrated parameters within a minute. 

We identified several model settings that calibrate well to the short maturity 
data, in particular a model setting with a local volatility that is piece-wise
linear in the benchmark forward rate together with an uncorrelated CIR 
stochastic variance and several model settings with a correlated lognormal 
stochastic volatility with quadratic drift (as introduced in \cite{LNSVSeppetal_2023})
combined with various local volatility forms, including the one that is 
linear in the benchmark forward rate. We prefer the lognormal stochastic volatility 
with quadratic drift since then we do not need to find the appropriate anchor 
strikes for the piece-wise linear local volatility by trial and error, 
instead of mostly automatically through some optimization. 
Benchmark forward rate formulations are preferred 
since they can be more easily fitted. 

Similar case and modeling studies can be performed with other model types 
and instruments. 
 
We see that the presented generic simulation, pricing, and calibration frameworks 
make such calibration and modelling studies feasible, efficient, and even easy. 
Based on their results, one can develop and fine tune appropriate 
production models for a variety of markets and instrument features. 

\newpage
{\bf Acknowledgments:}

The authors thank Vijayan Nair for his comments and suggestions regarding this research. 
Any opinions, findings and conclusions or recommendations expressed in this material are those of the
authors and do not necessarily reflect the views of Wells Fargo Bank, N.A., its parent
company, affiliates and subsidiaries.

\bibliographystyle{alpha}
\bibliography{../Bibliographies/cheyette}

\end{document}